\documentclass{LT23auth}
\usepackage{graphicx}

\begin{document}

\begin{frontmatter}
\title{ Relation between d-density wave of electron and staggered flux of spinon }
\author[address1]{Koichi Hamada\thanksref{thank1}},
\author[address1]{Daijiro Yoshioka}
\address[address1]{Department of Basic Science, University of Tokyo, Komaba 3-8-1, 
Meguro-ku, Tokyo 153-8902, Japan }
\thanks[thank1]{
E-mail: koichi@sola.c.u-tokyo.ac.jp}
\begin{abstract}

A $d_{x^2-y^2}$-density wave (ddw) order of electron in two-dimensional t-J model is analyzed 
in saddle point level using the U(1) slave boson formalism. 
We considered not only the staggered flux (s-flux) order of spinon but also the s-flux order of holon.
This analysis provides the relation between the s-flux order of spinon and the ddw order of electron. 
We discovered a new phase in the phase diagram.
In this phase, there is a s-flux order of spinon, but no ddw order of electron.
Our results are that 1) a region of electron ddw exists,
2) there is no coexistence of ddw and $d_{x^2-y^2}$-wave pairing (singlet-RVB) 
in all region of phase diagram, and that 
3) the ground state is a purely $d_{x^2-y^2}$ wave superconducting state.
\end{abstract}
\end{frontmatter}

 Recently, Chakravarty et al. \cite{CL01} proposed that the electron $d_{x^2-y^2}$-density 
wave (ddw) order exists in the pseudo gap region of high-$T_c$ superconductor.
The electron ddw state \cite{Nayak00} is the staggered flux (s-flux) state \cite{AM88,MA89} 
of electron coordinate.
In this state, $d_{x^2-y^2}$ wave (d-wave) gap exists, 
time-reversal-symmetry is broken and the `real' staggered current of the electron exists.
The order parameter of the ddw is 
$y_\mathrm{e}= -i \sum_{{\bf k} \sigma} ( \cos k_{x} -\cos k_{y} ) 
\langle c^{\dagger}_{{\bf k}\sigma} c_{{\bf k+Q} \sigma} \rangle $
, ${\bf Q}=(\pi,\pi)$.
However they didn't disscuss this scenario microscopically. 
There remains a question, `can the electron ddw phase exist in highly correlated system?'.

 The 2-dimensional t-J model is a promising model which includes highly correlated effects.
Many phases are proposed in this model \cite{AM88,MA89,Anderson87,Affleck88,Kotliar88,Zhang90,UL92}.
 Zhang \cite{Zhang90} analyzed the competition between s-flux order and $d_{x^2-y^2}$-wave 
pairing order at zero temparature by Gutzwillar approximation.
The s-flux state is unstable against infinitesimal d-wave pairing 
at finite doping.
 Ubbens and Lee \cite{UL92} analyzed  this model at finite temperature. 
The s-flux phase of the spinon exsists on the region 
where doping and temparature are both finite. 
The SU(2) s-flux state, in which the Fermi-surface of spinon are always points,  
is also analyzed in SU(2) slave boson model \cite{WenLee96,Lee98}.
However, there remains a question, 
`how do electrons behave in finite temperature s-flux phase?'.
The current of the spinon and the electron are not equivalent 
in finite tempereture s-flux phase 
bacause there is no Bose condensation of holon.
The region of finite temperature s-flux phase exists above 
the temperature of holon condensation.   
 
 In this paper, we revealed the relation between the ddw of electron and the s-flux of spinon.
We analyzed it in the 2-dimensional t-J model based on the U(1) slave boson formalism. 
We introduce order parameters
$ {\bar \chi_{ij}} =  \langle \sum_{\sigma} f_{i\sigma}^{\dag}f_{j\sigma} \rangle $,
$ {\bar \eta_{ij}} =  \langle f_{i\uparrow}f_{j\downarrow}-f_{i\downarrow}f_{j\uparrow} \rangle $,
$ {\bar {B}_{ij} } = \langle b^{\dag}_{i} b_{j} \rangle $ 
to decouple the Hamiltonian.
\begin{figure}[h]
\begin{center}\leavevmode
\includegraphics[width=0.7\linewidth]{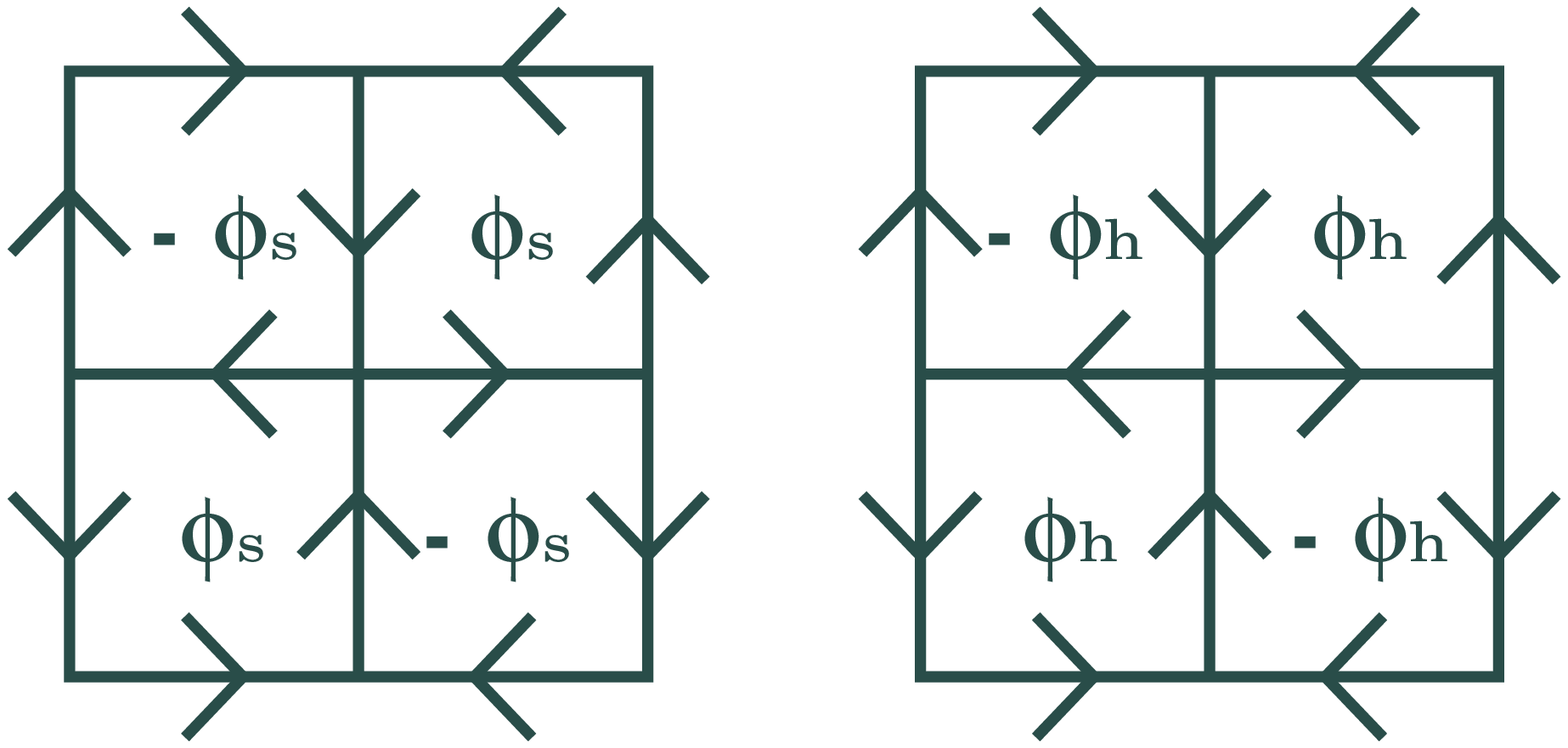}
\caption{ Staggered flux of spinon $\phi_\mathrm{s}$ and holon $\phi_\mathrm{h}$.}
\label{stag}\end{center}\end{figure}

We considered not only the staggered flux order of the spinon 
but also the staggerd flux order of the holon,
 ${\bar \chi}_{i+ {\hat x},i}
= x_\mathrm{s} + {\rm i}(-1)^{i} y_\mathrm{s} $
, 
 ${\bar \chi}_{i+ {\hat y},i}
= x_\mathrm{s} - {\rm i}(-1)^{i} y_\mathrm{s}$
, 
 ${\bar B}_{i+ {\hat x},i}
= x_\mathrm{h} + {\rm i}(-1)^{i} y_\mathrm{h}$
, 
 ${\bar B}_{i+ {\hat y},i}
= x_\mathrm{h} - {\rm i}(-1)^{i} y_\mathrm{h}$.
Here, ${\hat x}$ and ${\hat y}$ are unit vectors in the $x$ and $y$ direction,
$x_\mathrm{s}= \chi \cos(\phi_\mathrm{s}/4)$, $y_\mathrm{s}=\ \chi \sin(\phi_\mathrm{s}/4)$, 
$x_\mathrm{h}= B \cos(\phi_\mathrm{h}/4)$, $y_\mathrm{h}= B \sin(\phi_\mathrm{h}/4)$.
The order parameters $y_\mathrm{s}$ and $y_\mathrm{h}$ correspond to the ddw order 
parameter of spinon and holon, 
respectively.
For the pairing symmetry, 
we considered $d_{x^2-y^2}$, namely  
${\bar \eta}_{i+ {\hat x},i}= -\ {\bar \eta}_{i+ {\hat y},i}=\eta$.

This formalism is an extension of the study by Ubbens and Lee \cite{UL92}.
The $\sum_{\sigma}f_{j\sigma}^{\dag}f_{i\sigma}$ term  couples not only to $(3J/8)\chi_{ij}$ 
but also to $tB_{ij}$, i.e.
the spinon feels the spinon s-flux($\phi_\mathrm{s}$) and the holon s-flux($\phi_\mathrm{h}$). 
The expectation values of the holon, $B$ and $\phi_\mathrm{h}$, have finite value for 
the solution of self-consistency equations. 
Two advantages exist in our formalism;  
1) this is a new saddle point solution whose free enery is lower than the previous one, 
2) this solution provides the relation between the ddw of the electron and the s-flux of the spinon.
In this formalism, the hopping order parameter of the electron  is 
a product of the hopping order parameters of spinon and holon.
\begin{eqnarray}
\langle \sum_{\sigma} c_{i\sigma}^{\dag}c_{j\sigma} \rangle
=\langle \sum_{\sigma} f_{i\sigma}^{\dag} f_{j\sigma} \rangle \langle  b_{j}^{\dag} b_{i} \rangle
=\ {\bar \chi}_{ij} {\bar B}_{ij}^{*} 
\end{eqnarray}
The electron s-flux order parameter $\phi_\mathrm{e}$ and 
the electron ddw order parameter $y_\mathrm{e}$ are given by 
$\phi_\mathrm{e}= \phi_\mathrm{s} -\phi_\mathrm{h}$, 
and $y_\mathrm{e}=\chi B \sin ( \phi_\mathrm{e}/4 )$.
\begin{figure}[h]
\begin{center}\leavevmode
\includegraphics[width=0.83\linewidth]{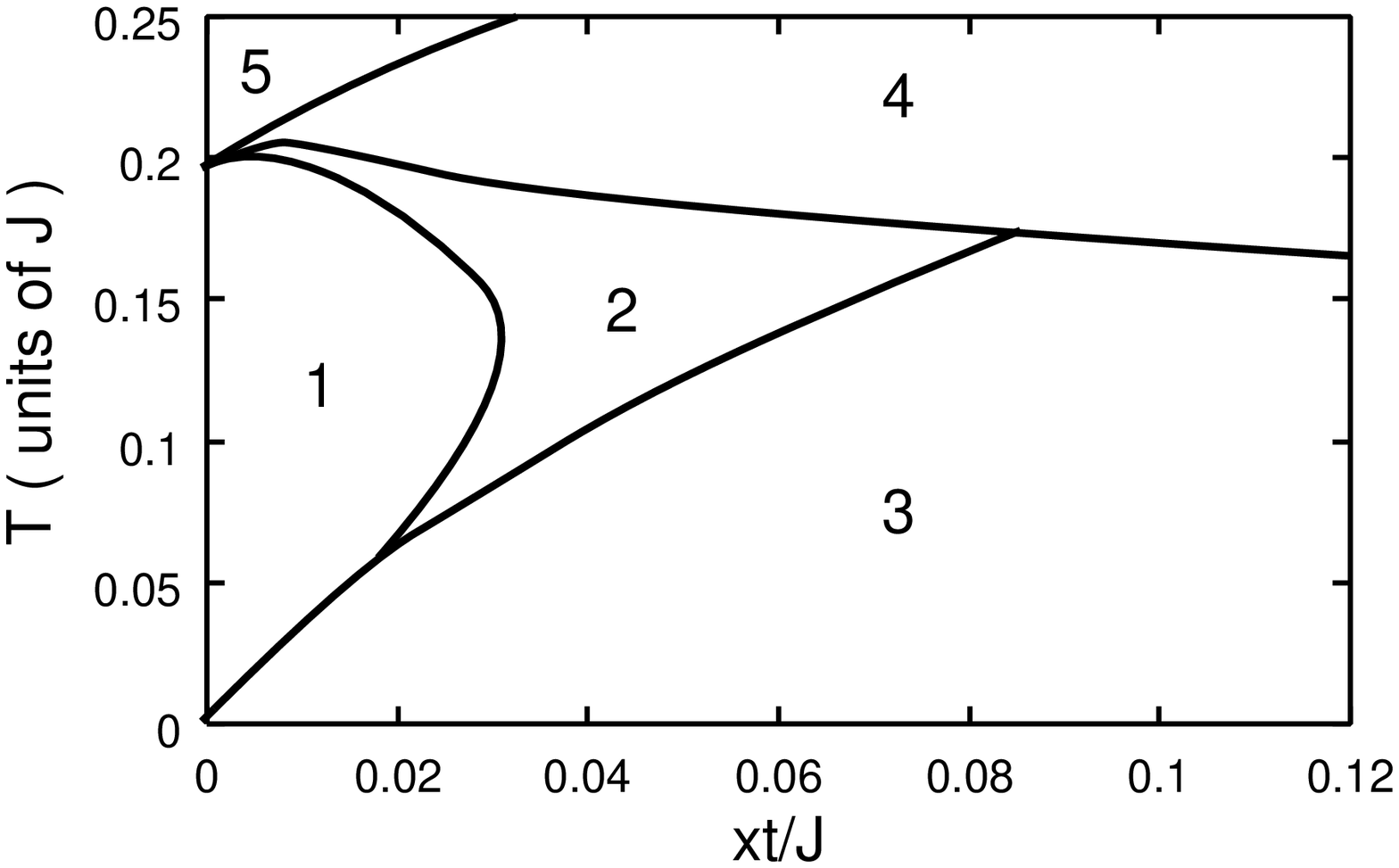}
\caption{ MF phase diagram for t/J=1.
The electron ddw order exists only in region 2. 
Here, x is hole concentration. 
The phase diagram for t/J=2 is quantitatively similar to the phase diagram for t/J =1.
With the boson order paramer $B_{ij}$, 
the $\pi$-flux phase and s-flux phase of spinon and holon 
extends to the higher-doped region compared to 
the previous work \cite{UL92}, where $B_{ij}$ were not considered.
}
\label{dgm}\end{center}\end{figure}

 We solved the self-consistency equations numerically,
and obtained the phase diagram(Fig.\ \ref{dgm}).
 At half-filling, the holon order parameter ${\bar B}_{ij}$ is zero 
and the degeneracy of spinon between the s-flux state and the d-wave pairing state exists 
due to the local SU(2) symmetry;
$\chi \ne 0, {\bar B}_{ij} =  0, y_\mathrm{s}^2 + \eta^2 = {\rm const} \ne 0$.
 In region 1, spinon s-flux exists but electron ddw doesn't exixt.
The spinon and holon state are $\pi$-flux order state respectively. 
In electron picture, the s-flux is canceled completely;  
$\chi \ne 0, B \ne 0, \phi_\mathrm{s} = \phi_\mathrm{h}= \pi, \eta = 0$.
The electron ddw order parameter, $y_\mathrm{e}=\chi B \sin\big( (\pi - \pi)/4 \big) =0$.
 The electron ddw order exists only in region 2.
The staggered current of electron can be observed 
experimentally.
The spinon s-flux $\phi_\mathrm{s}$ and holon s-flux $\phi_\mathrm{h}$ are not equal to $\pi$ or 0, 
and $\phi_\mathrm{s} \ne \phi_\mathrm{h}$;  
$\chi \ne 0, B \ne 0, \phi_\mathrm{s} \ne0, \phi_\mathrm{h} \ne 0, \eta = 0$, 
and $y_\mathrm{e}=\chi B \sin\big( (\phi_\mathrm{s} - \phi_\mathrm{h})/4 \big) \ne 0$.
 In region 3, $d_{x^2-y^2}$-wave pairing exists; 
$\chi \ne 0$ and $B \ne 0, \phi_\mathrm{s} = \phi_\mathrm{h} = 0, \eta \ne 0$.
and $y_\mathrm{e} = 0$.
 In region 4, there exists only uniform hopping order; 
$\chi \ne 0, B \ne 0, \phi_\mathrm{s} = \phi_\mathrm{h} = \eta = 0$, 
and $y_\mathrm{e} = 0$.
 In region 5, all order parameters are zero.
Spinon and holon cannot hop; 
$\chi = B = \phi_\mathrm{s} = \phi_\mathrm{h} = \eta = 0$.
and $y_\mathrm{e} = 0$.

 The transition between region 1 and region 2   
is a 2nd order transition in our theory.
(If one only focuses on spinon degree of freedom, 
this doesn't look like phase transition \cite{UL92,WenLee96,Lee98}.)
There exists an order parameter which charactrizes this transition. 
It is the electron ddw order parameter.

In conclusion we have found a possibility for the electron ddw phase 
in the t-J model at finite temperature, 
which does not coexists with the singlet-RVB state.
\begin{ack}
K.H. thanks A. Himeda, T. Koretsune, Y. Yanase, M. Ogata, Y. Ueno, K. Nomura, 
R. Shindou, S. Ryu, Y. Morita, J. Kishine, and N. Nagaosa 
for their useful discussions.  
Numerical computation in this work was partially carried out 
at the Yukawa Institute Computer Facility.
\end{ack}

%
%

\end{document}